\documentclass[11pt]{amsart}

\usepackage[dvips]{graphicx}
\usepackage{fancyhdr}

\usepackage{amssymb}
\usepackage[all]{xy}
\usepackage[colorlinks=true, urlcolor=rltblue, citecolor=drkgreen] {hyperref}
\usepackage{color}
\usepackage[mathscr]{eucal}
\usepackage{pdfsync}
\definecolor{rltblue}{rgb}{0,0,0.4}
\definecolor{drkgreen}{rgb}{0,0.4,0}

\newtheorem{thm}{Theorem}[section]

\theoremstyle{definition}

\theoremstyle{remark}

\newtheorem{historic}[thm]{Historic Remark}

\theoremstyle{plain}

\newcounter{contenumi}

\def\and{\mathrel{\&}}

\def\si{\sigma}
\def\b{\beta}

\def\b{\beta}

\def\H{\mathcal H}

\def\R{\mathtt{R}}

\def\RR{{\mathbb R}}

\DeclareMathOperator{\gam}{Gamma}

\def\Ssf{\mathsf S}
\def\Esf{\mathsf E}
\def\Isf{\mathsf I}
\def\Rsf{\mathsf R}

\def\Si{{\mathsf S}}
\def\Sb{\bar{{\mathsf S}}}
\def\Sbb{\overline{\overline {\mathsf S}}}
\def\qbb{\overline{\overline {\mathsf q}}}
\def\Ri{{\mathsf R}}
\def\Rb{\bar{{\mathsf R}}}
\def\Rbb{\overline{\overline {\mathsf R}}}
\def\Ii{{\mathsf I}}
\def\Ib{\bar{{\mathsf I}}}

\def\Ei{{\mathsf E}}
\def\Eb{\bar{{\mathsf E}}}

\def\R{{\mathscr R}}
\def\Reff{\R_{\text{eff}}}
\def\Ro{\R_{0}}
\def\Rt{\R_{\text{t}}}

\def\infe{{\mathsf f}}
\def\inft{{\mathsf t}}

\title{Herd immunity under individual variation and reinfection}

\author{Antonio Montalb\'an}
\address{Department of Mathematics\\
University of California, Berkeley\\
 USA}
\email{antonio@math.berkeley.edu}
\urladdr{\href{http://www.math.berkeley.edu/~antonio/index.html}{www.math.berkeley.edu/$\sim$antonio}}

\author{Rodrigo M. Corder}
\address{Department of Epidemiology and Biostatistics\\
University of California, Berkeley School of Public Health, Berkeley\\
USA}
\email{rodrigo.corder@berkeley.edu}

\author{M. Gabriela M. Gomes}
\address{Department of Mathematics and Statistics, University of Strathclyde, Glasgow, UK}
\email{Gabriela.Gomes@strath.ac.uk}
\urladdr{\href{https://www.strath.ac.uk/staff/gomesgabrieladr/}{www.strath.ac.uk/staff/gomesgabrieladr}}

\begin{document}

\maketitle


%

\begin{abstract}
We study a susceptible-exposed-infected-recovered (SEIR) model considered by Gomes et al. \cite{Gomes2022,Aguas2020} where individuals are assumed to differ in their susceptibility or exposure to infection. 
Under this heterogeneity assumption, epidemic growth is effectively suppressed when the percentage of the population having acquired immunity surpasses a critical level - the herd immunity threshold - that is lower than in homogeneous populations.
We derive explicit formulas to calculate herd immunity thresholds and stable configurations, especially when susceptibility or exposure are gamma distributed, and explore extensions of the model. 
\end{abstract}

\section{Introduction}

Understanding and predicting the dynamics and control of infectious diseases relies on representative models, whether conceptual or mathematical. Mathematical modelling was established in infectious diseases over a century ago, with the seminal works of Ross and Hudson \cite{Ross,RH}, Kermack and McKendrick \cite{KM} and others. Propelled by the discovery of aetiological agents for infectious diseases, and the germ theory, models have focused on the complexities of pathogen transmission and evolution \cite{Heesterbeek}. It has recurrently been noted for over a century, however, that these models tend to overpredict transmission potential and overestimate the impact of control measures which may be explained by limitations in capturing the effects of heterogeneity \cite{KM,McKendrick,Gart68,Gart71,Ball,AMMJ,P-S&V,Miller,Gomes2022}. 

Here we analyze a set of susceptible-exposed-infected-recovered (SEIR) models presented in \cite{Gomes2022,Aguas2020} where each of the compartments $\Ssf$, $\Esf$, $\Isf$ and $\Rsf$ is expanded into continuum many compartments $S(x)$, $E(x)$, $I(x)$ and $R(x)$, where $x\in \RR^+$ is a trait that varies among individuals.
Specifically we model a situation where each individual has a level of susceptibility or exposure (connectivity) $x$, starting in compartment $S(x)$ and staying within the compartments $S(x)$, $E(x)$, $I(x)$ and $R(x)$ the whole time.
This individual may infect or be infected by others irrespective of their trait value $x$ assuming random mixing \cite{AM1991,DHB}.
We will consider two types of models:

A {\bf variable susceptibility} case where the susceptibility of an individual at level $x$ is proportional to $x$, or, in other words, if we compare an individual at level $x$ and an individual at level $y$, the one at level $x$ is $x/y$ times more likely to get infected than the one with susceptibility $y$. 
We may interpret this as variation in biological susceptibility which may be due to genetics, epigenetics or life history. 

A {\bf variable connectivity} case where the propensities for an individual at level $x$ to acquire infection and transmit to others are both proportional to $x$, or, in other words,  if we compare an individual at level $x$ and an individual at level $y$, the one at level $x$ is $x/y$ times more likely to get infected than the one in level $y$ and also $x/y$ times more likely to infect someone else once infected. 
This is interpreted as individuals with many contacts being both more likely to get infected and to infect others.

For each $x$, we have a system of the form:
\[
\xymatrix{
\fbox{$S(x)$}   \ar[r]^{x\lambda}    &     \fbox{$E(x)$}   \ar[r]^{\delta}   &  \fbox{$I(x)$}  \ar[r]^{\gamma  }  &  \fbox{$R(x)$}
}
\]
where $\lambda$ is the force of infection which is formulated differently in the variable susceptibility or the variable connectivity cases:
\begin{eqnarray*}
\mbox{Variable susceptibility:} 	&\quad& 	\lambda = \b \int I(x) \ dx,  \\
\mbox{Variable connectivity:} 	&\quad& 	\lambda = \b \int x \ I(x) \ dx.  
\end{eqnarray*}
Note that $\lambda$ varies with time, as it depends on the time-dependent infected population.   

The dynamics of compartments $S(x)$, $E(x)$, $I(x)$ and $R(x)$ are governed by the infinite system of ordinary differential equations:
\begin{eqnarray}
\frac{dS(x)}{dt} &=& - \lambda\ x\ S(x),   \label{introSgen}\\ 
\frac{dE(x)}{dt}  &=&  \lambda\ x\ S(x) -\delta\ E(x),  \label{introEgen}\\ 
\frac{dI(x)}{dt} &=&  \delta\ E(x) - \gamma\ I(x),   \label{introIgen}\\
\frac{dR(x)}{dt} &=&  \gamma\ I(x). \label{introRgen}
\end{eqnarray}

We assume that the system has been scaled such that the total population is 1.
The initial conditions for variables $S(x,t)$, $E(x,t)$, $I(x,t)$ and $R(x,t)$, satisfy $S(x,0)=(1-\epsilon)\ q(x)$, $E(x,0)=\epsilon\ q(x)$ and $I(x,0)=R(x,0)=0$, where $0<\epsilon\ll 1$ is a small scalar to seed the epidemic, and $q(x)$ is a probability density function with mean 1 and coefficient of variation $\nu$:
\begin{eqnarray}
\int x q(x)\ dx = 1 \qquad &\mbox{and}& \qquad \sqrt{\int (x-1)^2 q(x)\ dx} = \nu.
\end{eqnarray}

We use $\Si(t)$ to denote the integral over all susceptibility levels, $S(x,t)$, for $x\in\RR^+$. 
We thus have $\Si(t)=\int_0^{+\infty} S(x,t) dx$.
Same with $\Ei(t)$, $\Ii(t)$ and $\Ri(t)$.

We will use the first three moments of $S(x,t)$, that we denote $\Si(t)$, $\Sb(t)$ and $\Sbb(t)$:

\begin{eqnarray}
\Si(t) = \int S(x,t)\ dx,
\quad\mbox{}\quad
\Sb(t) = \int x S(x,t)\ dx
\quad\mbox{and}\quad
\Sbb(t) = \int x^2 S(x,t)\ dx.
\end{eqnarray}

When infection is absent ($\epsilon=0$), we have $\Si(0)=1$, $\Sb(0)=1$ and $\Sbb(0) =  1 + \nu^2$. 
But note that $S(x,t)$ is not a probability density function for $\epsilon >0$ as $\Si(t)$ becomes less than $1$.
The quotient $S(x,t)/\Si(t)$ as a function of $x$ for fixed $t$ will be a probability density function for $\epsilon >0$ and all $t$ with first and second moments $\Sb(t)/\Si(t)$ and $\Sbb(t)/\Si(t)$ which decrease over time.
When the initial configuration $q(x)$ is a gamma distribution, all the distributions $S(x,t)/\Si(t)$ are also gamma with the same coefficient of variation $\nu$ but with lower mean (see Appendix \ref{app:gamma}), an argument which enables mathematical derivations to advance further when traits are assumed to be gamma distributed \cite{Novozhilov}. 

Similarly, we define the moments $\Ri(t)$, $\Rb(t)$ and $\Rbb(t)$ for the recovered compartment, and the same with $\Esf$ and $\Isf$.
Notice for instance that $\lambda(t)$ is written as $\beta\ \Isf(t)$ and $\beta\ \Ib(t)$ in the variable susceptibility and variable connectivity cases, respectively. 

Here we describe key epidemiological quantities when system of equations (Eqs. \ref{introSgen}-\ref{introRgen}) is adopted. The basic reproduction number $\Ro$ is the average number of secondary infections generated by an infected individual in a totally susceptible population. It depends on characteristics of both the pathogen and the host population. When this number is below 1 no epidemics are expected. When $\Ro$ is above 1, however, the introduction of infection in a virgin population is expected to generate an epidemic. This is followed by almost exponential growth in cumulative infections which decelerates gradually as susceptibles are depleted. The effective reproduction number $\Reff$ is a time-dependent quantity loosely defined as the number of secondary infections generated by a typical infected individual when the susceptibility of the population is as at time $t$. $\Reff$ coincides with $\Ro$ at the beginning of an epidemic (when the population is totally susceptible) but declines as individuals are removed from the susceptible pool by infection and immunity. As $\Reff$ crosses 1 towards lower values, the epidemic subsides and future reintroductions of infection are not expected to generate new outbreaks as long as population immunity is maintained.

We derive formulas for the effective reproduction number $\Reff$ and the herd immunity threshold $\H$ in terms of moments $\Si$, $\Sb$ and $\Sbb$, of the susceptible population.
When $q(x)$ is a gamma distribution, $\Sb$ and $\Sbb$ can be formulated in terms of $\Si$ and we get an exact formula for $\H$ in terms only of the basic reproduction number $\Ro$ and the coefficient of variation $\nu$. In this case we can also reduce the infinite system (Eqs. \ref{introSgen}-\ref{introRgen}) to a finite system of ordinary differential equations in $\Ssf$, $\Esf$, $\Isf$ and $\Rsf$ with nonlinear transmission (exactly when the variable trait is susceptibility and approximately in the case of variable connectivity). In the case of variable connectivity, we provide an exact derivation of a finite system in the variables $\Sb$, $\Eb$, $\Ib$ and $\Rb$.

\

In the {\bf variable susceptibility} case we will get that:
\[
\Ro = \frac{\beta}{\gamma}
\quad\mbox{and}\quad
{\Reff} = \frac{\b}{\gamma}\ \Sb,
\]
and consequently, ${\Reff} =\Ro\ \Sb$. This implies that the population is above the herd immunity threshold when $\Sb < 1/\Ro$.
When we assume that $q(x)$ is a gamma distribution, the proportion of individuals that have been infected by the time the herd immunity threshold is reached is deduced as:
\begin{eqnarray} \label{HITs}
\H = 1 - \Ro^{-\frac{1}{1+\nu^2}},
\end{eqnarray} 
and system (Eqs. \ref{introSgen}-\ref{introRgen}) can be reduced to:
\begin{eqnarray}
\frac{d\Ssf}{dt} &=& - \beta\ \Isf\ \Ssf^{1+\nu^2},   \label{introSsus}\\ 
\frac{d\Esf}{dt}  &=&  \beta\ \Isf\ \Ssf^{1+\nu^2} -\delta\ \Esf,  \label{introEsus}\\ 
\frac{d\Isf}{dt}  &=&  \delta\ \Esf - \gamma\ \Isf,   \label{introIsus}\\
\frac{d\Rsf}{dt}  &=&  \gamma\ \Isf. \label{introRsus}
\end{eqnarray}
These equations are exact.

\

In the {\bf variable connectivity} case we will get that:
\[
\Ro = (1+\nu^2)\ \frac{\beta}{\gamma}
\quad\mbox{and}\quad
{\Reff} =  \frac{\b}{\gamma}\ \Sbb,
\]
and consequently, ${\Reff} =\Ro/(1+\nu^2)\ \Sbb$. This implies that the population is above the herd immunity threshold when $\Sbb < (1+\nu^2)/\Ro$.
When we assume that $q(x)$ is a gamma distribution, the proportion of individuals that have been infected by the time the herd immunity threshold is reached is deduced as:
\begin{eqnarray} \label{HITc}
\H = 1 - \Ro^{-\frac{1}{1+2\nu^2}}.
\end{eqnarray} 
In (Eqs. \ref{Scon0}-\ref{Rcon0}), we derive a closed system for the variable connectivity model on the variables $\Sb$, $\Eb$, $\Ib$ and $\Rb$. But this does not directly allow the model to be fitted to incidence data provided by routine surveillance. For finding equations that determine the system only on the variables  $\Ssf$,  $\Esf$,  $\Isf$ and  $\Rsf$, in the variable connectivity case we need to make an approximation as detailed in Appendix \ref{app:connectivity}. The resulting system (Eqs. \ref{Ssus}-\ref{Rsus}) is shown to approximate the original (Eqs. \ref{introSgen}-\ref{introRgen}) when the infectious period is small as in acute infectious diseases.

\

In Figure 1 we provide graphical representations for the $\H$ formulas corresponding to the basic model introduced in this section (variable susceptibility in green and variable connectivity in blue), showing a monotonic decrease as the coefficient of variation $\nu$ increases (\cite{Gomes2022}).

\begin{figure}
 \centerline{\includegraphics[clip, trim=4cm 15.5cm 7cm 7.5cm, width=.5\textwidth]{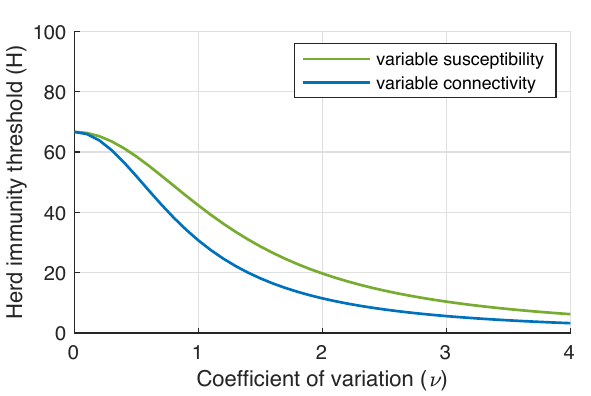}}
\caption{Herd immunity threshold. Curves generated using formulas (Eq. \ref{HITs}) for gamma distributed susceptibility and (Eq. \ref{HITc}) for gamma distributed connectivity, with $\Ro=3$.}\label{fig:HIT}
\end{figure}

\


In Sections \ref{Reff} and \ref{HIT} we provide general derivations for effective reproduction numbers and herd immunity thresholds, while Section \ref{gamma} is focussed on special cases when traits $x$ are gamma distributed. Towards the end of the paper, we analyze two extensions of the basic model.
In Section \ref{sec:reinfection}, we consider a model with reinfection where immunity after recovery is not fully protective but only partially.
In Section \ref{sec: carrier}, we consider a model with a carrier state, which \cite{Gomes2022,Aguas2020} apply to the coronavirus disease (COVID-19) pandemic.
There, the exposed compartments are not simply latent but a carrier state where individuals are infectious but to a lesser degree than individuals in the fully infectious compartment. 
In each case we derive formulas for herd immunity thresholds especially when the initial trait distribution is gamma.

\section{Effective reproduction number} \label{Reff}

The {\em effective reproduction number} at time $t$ is defined as the number of secondary infections caused by a typical infected individual over their entire infectious period in an idealized situation, as described below.

Before proceeding with its derivation we need to make two considerations. The first concerns the evolving susceptible pool. We assume that during the period in which an individual is contagious, the density of susceptibles is frozen in time. This means that we disregard the fact that since the susceptible population declines, this individual infects less at the end than at the beginning of their infection. In acute infections, the decline in the susceptible population is usually slow compared to the rate of recovery from infection so, in practice, the impact of the assumption is negligible. Moreover, when $\Reff$ is used to analyze stable configurations, such as in the derivation of herd immunity thresholds, the assumption holds and hence has no effect on the results. This consideration pertains to both variable susceptibility and variable connectivity models.

The second concerns the infectivity profile of the infected population at time $t$.  We define:
\begin{quote} 
 $\Reff$ at time $t$ as the average number of secondary infections generated by an individual who {\em becomes infected} at time $t$.  
This average is taken over the pool of individuals that go from $\Ssf$ to $\Esf$ at time $t$. 
\end{quote}
When $\Reff<1$, infection is not expected to invade an infection-free population.
Further details on this concept are discussed in Appendix \ref{app:connectivity}. 
In the remaining of this section we derive explicit formulas for $\Reff$.

\

First, the {\bf variable susceptibility} case:
Consider an individual who gets infected (more precisely, exposed and consequently infected) at time $t$ (i.e., moves from $\Ssf$ to $\Esf$ at time $t$).
This individual will eventually move to $\Isf$ and spend on average $1/\gamma$ days there.
While in $\Isf$, the individual will infect an average of $\beta \int y\ S(y,t)\ dy$ others per day.
We thus get:
\begin{eqnarray}
{\Reff} (t) &=&  \frac{\beta}{\gamma}\  \int y\ S(y,t)\ dy  =  \frac{\beta}{\gamma}\ \Sb(t) . \label{Reff_sus}
\end{eqnarray}

In particular, we get:
\begin{eqnarray}
\Ro&=&\frac{\beta}{\gamma}, \label{R0_sus}
\end{eqnarray}
and consequently:
\begin{eqnarray}
{\Reff}(t) &=& \Ro\  \Sb(t). \label{ReffR0_sus}
\end{eqnarray}

Second, the {\bf variable connectivity} case:
Consider again an individual who gets infected at time $t$.
It now matters what trait value $x$ this individual has because it determines how many others they will infect.  

Let $p(x,t)$ be the density function measuring the probability at time $t$ that this individual has connectivity level $x$.
The probability of becoming infected (i.e., of entering the $\Esf$ compartment) is $x \lambda(t)$. 
Thus, the value of $p(x,t)$ is proportional to $x S(x,t)$: 
\begin{eqnarray*}
p(x,t) &=& x\ \frac{S(x,t)}{\Sb(t)}.
\end{eqnarray*}
As above, an individual who enters $\Esf$ will eventually move to $\Isf$, spend on average $1/\gamma$ days there, and infect an average of $\beta \int y\ S(y,t)\ dy$ others per day.
We thus get:
\begin{eqnarray}
{\Reff}(t) &=& \int \frac{\beta}{\gamma}\ \Big(\int y\ S(y,t)\ dy\Big)\ x \ p(x,t)\ dx =
	\frac{\beta}{\gamma}\ \int \Sb(t)\ \frac{x^2\ S(x,t)}{\Sb(t)}\  dx =
        \frac{\beta}{\gamma}\ \Sbb(t).
\label{Reff_con}
\end{eqnarray}

In particular, we get:
\begin{eqnarray}
\Ro&=&   \qbb\ \frac{\beta}{\gamma} = (1+\nu^2)\ \frac{\beta}{\gamma}, \label{R0_con}
\end{eqnarray}
and consequently:
\begin{eqnarray}
{\Reff}(t) &=& \frac{\Ro}{1+\nu^2}\ \Sbb(t). \label{ReffR0_con}
\end{eqnarray}

Expressions for $\Ro$, such as (Eq. \ref{R0_sus}) and (Eq. \ref{R0_con}), have been known for decades \cite{AM1991,DHB,Woolhouse}. Worth highlighting, however, is that disproportionately less attention has been given to variable susceptibility than to variable connectivity due to the coefficient of variation $\nu$ not affecting the formula explicitly in the former case but only in the latter. It will be important to realise, however, that variation in susceptibility is just as impactful when we consider quantities such as herd immunity thresholds and  inferences of $\Ro$ from observational data \cite{Gomes2022,Aguas2020}.

\section{Herd immunity threshold} \label{HIT}

Suppose we have a population with no infected individuals, so that all individuals are either susceptible or recovered. The population is said to be at or above the herd immunity threshold for a pathogen if its susceptibility profile to that pathogen is such that a new introduction of infection (i.e., a small increase in $\Esf$ or $\Isf$) does not trigger an outbreak.
By inspection on the differential equations (Eqs. \ref{introSgen}-\ref{introRgen}), we see that a configuration with $E(x)=I(x)=0$ satisfies this condition if and only if
\[
{\Reff} \leq 1.
\]

In the variable susceptibility case it is equivalent to formulate the herd immunity threshold in terms of suppression of future outbreaks (as adopted here) or in terms of an unmitigated epidemic passing its peak infection prevalence. With variable connectivity, however, this equivalence does not hold as explained in Section \ref{Reff}.

A configuration with no infected individuals is then said to be at the herd immunity threshold if  and only if ${\Reff}=1$.
In SEIR models with no individual variation, configurations with no infected individuals are determined by the values of $\Ssf$ and $\Rsf=1-\Ssf$, and the herd immunity threshold is defined as the value of $1-\Ssf$ at the unique configuration with ${\Reff}=1$.
This value is well-known to be equal to $1-1/\Ro$ \cite{AM1991,DHB}.
With individual variation in susceptibility or exposure to infection, however, there are many configurations which satisfy ${\Reff}=1$.
One such configuration is given by $S(x) = q(x)/\Ro$ for all $x$. 
This could be obtained, for instance, by vaccinating a proportion $1-1/\Ro$ of the total population randomly without taking into account susceptibility or exposure levels \cite{Fine}.
When immunity is acquired naturally, however, individuals with higher susceptibility or exposure tend to be infected earlier and the herd immunity threshold is reached before the susceptible population is as low as $1/\Ro$ of the total. For now, let us retain that for the basic heterogeneous models considered here the herd immunity threshold is reached when $\Sb=1/\Ro$ in the {\bf variable susceptibility} case (Eq. \ref{ReffR0_sus}) and $\Sbb/(1+\nu^2)= 1/\Ro$ with {\bf variable connectivity} (Eq. \ref{ReffR0_con}). 

Next, we will see how we can derive $\Sb(t)$ and $\Sbb(t)$ under the assumption that the initial distribution is gamma.

\section{Case of the gamma distribution} \label{gamma}

Here we study how the distribution of the trait $x$ within the susceptible compartment evolves generally, and then specify to the case where individual variation in susceptibility or connectivity is gamma distributed.  We also refer to related work in \cite{Neipel,Novozhilov}.

\subsection{Evolution of the susceptible compartment}

From the SEIR equation for $d S(x,t)/dt$ we get:
\begin{eqnarray*}
\frac{1}{S(x,t)} \frac{d S(x,t)}{d t}  &=& - x  \ \lambda(t).
\end{eqnarray*}
Integrating with respect to $t$ we get: 
\begin{equation} \label{eq: stx}
S(x,t) = q(x) \ e^{-x\ k_t}
\quad
\mbox{where}
\quad
k_t= \int_0^t \lambda(u)\ du.
\end{equation}

This holds in both variable susceptibility and variable connectivity models (with different values for $k_t$). It also holds in the cases with reinfection and with carrier state considered in Sections \ref{sec:reinfection} and \ref{sec: carrier}.

%
%

\subsection{Gamma distributed traits}

The key observation here is that since $S(x,t) = q(x) \ e^{-x\ k_t}$, we have that $S(x,t)/\Si(t)$ remains a gamma distribution at all values of $t$.
This enables the derivations in Appendix \ref{app:gamma} of explicit formulas for the moments $\Sb$ (Eq. \ref{Sb_gamma}) and $\Sbb$ (Eq. \ref{Sbb_gamma}) in terms of $\Si$ and the shape parameter $\alpha$, when susceptibility or connectivity are gamma distributed.

Using that the coefficient of variation is $\nu= 1/\sqrt{\alpha}$ we rewrite the respective formulas as:
\begin{eqnarray}
\Sb(t) &=& \Si(t)^{1+\nu^2},  \label{Sbar} \\
\Sbb(t) &=& (1+\nu^2)\ \Si(t)^{1+2 \nu^2}. \label{Sbarbar}
\end{eqnarray}

\

To derive the reduced system in the {\bf variable susceptibility} case (Eqs. \ref{introSsus}-\ref{introRsus}) we integrate the equations in (Eqs. \ref{introSgen}-\ref{introRgen}) over the susceptibility domain and apply (Eqs. \ref{Reff_sus} and \ref{Sbar}) to get:
\begin{eqnarray}
\frac{d\Ssf}{dt} &=& - \beta\ \Isf\ \Sb = - \beta\ \Isf\ \Ssf^{1+\nu^2},  \label{Ssus}\\ 
\frac{d\Esf}{dt}  &=& \beta\ \Isf\ \Sb -\delta\ \Esf = \beta\ \Isf\ \Ssf^{1+\nu^2} -\delta\ \Esf, \label{Esus}\\ 
\frac{d\Isf}{dt}  &=&  \delta\ \Esf - \gamma\ \Isf, \label{Isus}\\
\frac{d\Rsf}{dt} &=&  \gamma\ \Isf. \label{Rsus}
\end{eqnarray}
This closed system in $\Ssf$, $\Esf$, $\Isf$ and $\Rsf$ has been used to fit epidemic curves of COVID-19 \cite{Gomes2022,Aguas2020}.
Recalling that the herd immunity threshold $\H$ is $1-\Si(t)$ at the time $t$ when ${\Reff}=1$ and that ${\Reff}(t) =  \Ro\ \Sb(t)$ (Eq. \ref{ReffR0_sus}), we get that $\Ro^{-1} = \Sb(t) = \Si(t)^{1+\nu^2}$ at the herd immunity threshold when susceptibility is gamma distributed. Hence: 
\begin{eqnarray}
\H &=& 1 - \Ro^{-\frac{1}{1+\nu^2}}. \label{H0_hetsus}
\end{eqnarray}
\

In the {\bf variable connectivity} case we have that ${\Reff}(t) = \Ro/(1+\nu^2)\ \Sbb(t)$  (Eq. \ref{ReffR0_con}).
Thus, when ${\Reff}=1$, we get $\Ro^{-1} = \Sbb(t)/(1+\nu^2) = \Si(t)^{1+2\nu^2}$ when connectivity is gamma distributed. Hence: 
\begin{eqnarray}
\H &=& 1 - \Ro^{-\frac{1}{1+ 2 \nu^2}}. \label{H0_hetcon}
\end{eqnarray}

Multiplying the original (Eqs. \ref{introSgen}-\ref{introRgen}) by $x$, integrating over the connectivity domain and applying (Eqs. \ref{Reff_con} and \ref{Sbarbar}) we get:
\begin{eqnarray}
\frac{d\Sb}{dt} &=& - \beta\ \Ib\ \Sbb = - (1+\nu^2)\ \beta\ \Ib\ \Sb^\frac{1+2\nu^2}{1+\nu^2} , \label{Scon0}\\ 
\frac{d\Eb}{dt} &=& \beta\ \Ib\  \Sbb -\delta\ \Eb = (1+\nu^2)\ \beta\ \Ib\ \Sb^\frac{1+2\nu^2}{1+\nu^2}  -\delta\ \Eb, \label{Econ0}\\ 
\frac{d\Ib}{dt}  &= & \delta\ \Eb - \gamma\ \Ib, \label{Icon0}\\
\frac{d\Rb}{dt}  &= & \gamma\ \Ib. \label{Rcon0}
\end{eqnarray}

Mathematically this is a tractable closed system in $\Sb$, $\Eb$, $\Ib$ and $\Rb$. However, these variables are not convenient for practical data fitting and parameter estimation. In Appendix \ref{app:connectivity} we propose an approximation in the variables $\Ssf$, $\Esf$, $\Isf$ and $\Rsf$.

\section{Model with reinfection} \label{sec:reinfection}

Here we consider an extension of the model considering that immunity after recovery is not fully protective, but only partially.
A factor $\si$, with $0\leq \si \leq 1$, is added to represent the quotient of the probability of getting reinfected after recovery over the probability of getting infected while fully susceptible. 

The model is represented diagrammatically as:
\[
\xymatrix{
\fbox{$S(x)$}   \ar[r]^{x\lambda}    &     \fbox{$E(x)$}   \ar[r]^{\delta}   &  \fbox{$I(x)$}  \ar[r]^{\gamma  }  &  \fbox{$R(x)$} \ar@/^2pc/[ll]^{\si x \lambda}
}
\]
with $\lambda$ as in the basic models (without reinfection) studied above.
The extended model is given by the equations:
\begin{eqnarray}
\frac{d S(x)}{dt} &=& - \lambda\  x\ S(x),  \label{Sreinf} \\ 
\frac{d E(x)}{dt}  &=&  \lambda\ x\ (S(x)+\si\ R(x)) -\delta\ E(x),   \label{Ereinf} \\ 
\frac{d I(x)}{dt}  &=&  \delta\ E(x) - \gamma\ I(x),    \label{Ireinf} \\
\frac{d R(x)}{dt}  &=&  \gamma\ I(x) - \si\  \lambda\ x\ R(x).  \label{Rreinf}
\end{eqnarray}

The system exhibits newer dynamics in comparison with the basic case.
Depending on whether $\si$ is below or above $1/\Ro$ (known as the reinfection threshold \cite{Gomes2004,Gomes2016}) we get that either the disease dies out after a while and a certain proportion of the population never gets infected, or continues endemically and every individual is eventually infected and reinfected repeatedly.

\subsection{Effective reproduction number}

The basic reproduction number is calculated exactly, as in the absence of reinfection, but the effective reproduction now depends not only on the distribution of $S(x,t)$, but also on the distribution of $R(x,t)$.
When we consider configurations with no infected individuals, we will have that $R(x,t)=q(x)-S(x,t)$ and will be able to express ${\Reff}(t)$ in terms of $S(x,t)$ only. 

The formulas for the effective reproduction number ${\Reff}$ at time $t$ are: 

\begin{itemize}
\item ${\Reff}(t) =  (\beta /\gamma)\ (\Sb(t)+ \si \Rb(t)) $ in the variable susceptibility case, and
\item ${\Reff}(t) = (\beta/\gamma)\ (\Sbb(t)+ \si \Rbb(t)) $ in the variable connectivity case.
\end{itemize}

The derivation of these formulas is essentially as the derivations in Section \ref{Reff}, with two differences.
First, each individual with trait value $x$ infects $\beta(\Sb(t)+ \si \Rb(t))$ or  $ \beta x(\Sb(t)+ \si \Rb(t))$ others per day spent in $\Isf$, in the respective cases, instead of $\beta\Sb(t)$ or $\beta x\Sb(t)$.
Second, when we consider an individual that gets infected in the variable connectivity case, the probability that this individual has trait value $x$ is proportional to $x(S(x,t)+\si R(x,t))$ instead of $xS(x,t)$.

\subsection{Herd immunity threshold}

Recall that a configuration with no infected individuals is at or above the herd immunity threshold if and only if ${\Reff} \leq 1.$
Assuming that no one is infected, that is $R(x)=q(x)-S(x)$, we get $\Rb=1-\Sb$ and $\Rbb=\qbb-\Sbb$, where $\qbb=\int x^2 q(x)\ dx = 1+\nu^2$.
We can then understand the configurations at the herd immunity threshold in terms of $\Sb$ and $\Sbb$.

In the {\bf variable susceptibility} case a configuration with no infected individuals is at the herd immunity threshold if and only if $\Sb +\si (1-\Sb)=1/\Ro$, and hence:
\[
\Sb = \frac{\Ro^{-1}-\si}{1-\si}.
\]

In the {\bf variable connectivity} case a configuration with no infected individuals is at the herd immunity threshold if and only if $(\Sbb +\si (\qbb-\Sbb))/(1+\nu^2)=1/\Ro$, and hence:
\[
\Sbb = (1+\nu^2)\ \frac{\Ro^{-1} - \si}{1-\si}.
\]


\subsection{Reinfection threshold}

The formulas above require:
\[
\si<\Ro^{-1}.
\] 
That is, the reinfection factor  $\si$ has to be below $\Ro^{-1}$, a critical value known as the reinfection threshold \cite{Gomes2004,Gomes2016}. If this is verified, then all configurations with no infected individuals and satisfying the conditions above (either $\Sb = (\Ro^{-1}-\si)/(1-\si)$ or $\Sbb = (1+\nu^2) (\Ro^{-1} - \si)/(1-\si)$ depending on the case) are herd immunity threshold configurations in the sense that when susceptibility is at that level or lower, infection reintroductions will not trigger new outbreaks as ${\Reff}$ will not increase above 1. 

If the reinfection factor is so high that the population is above the reinfection threshold, ${\Reff}$ will be greater than 1 in any such configurations, so there will not be any configuration with no infected individuals which is at the herd immunity threshold.
This implies that there will always be a portion of the population infected, and hence that the population of susceptible individuals will eventually be completely depleted. The infection becomes endemic.
The equilibrium configuration will now have $S(x)=0$ for all $x$. In these situations the level of endemicity will depend on how much resistance the population is able to mount and maintain.


\subsection{Case of the gamma distribution}

Recall from Section \ref{HIT} that, when the initial distribution $q(x)$ is gamma, we get that $S(x,t)/\Si(t)$ remains a gamma distribution for all $t$, and that $\Sb(t)=\Si(t)^{1+\nu^2}$ and $\Sbb(t)=(1+\nu^2)\ \Si(t)^{1+2\nu^2}$.
We can then obtain the values of $\Si(t)$ at the time when the herd immunity threshold is reached, and then calculate $\H$ as $1-\Si(t)$ for that particular $t$.

In the {\bf variable susceptibility} case we get:
\begin{eqnarray}
\H = 1 - \left(\frac{\Ro^{-1}-\si}{1-\si}\right)^{\frac{1}{1+\nu^2}}. \label{H0si_hetsus}
\end{eqnarray}

In the {\bf variable connectivity} case we get:
\begin{eqnarray}
\H = 1 - \left(\frac{\Ro^{-1}-\si}{1-\si}\right)^{\frac{1}{1+ 2 \nu^2}}. \label{H0si_hetcon}
\end{eqnarray}

\begin{figure}
 \centerline{\includegraphics[clip, trim=0.5cm 13.5cm 0.5cm 9cm, width=1\textwidth]{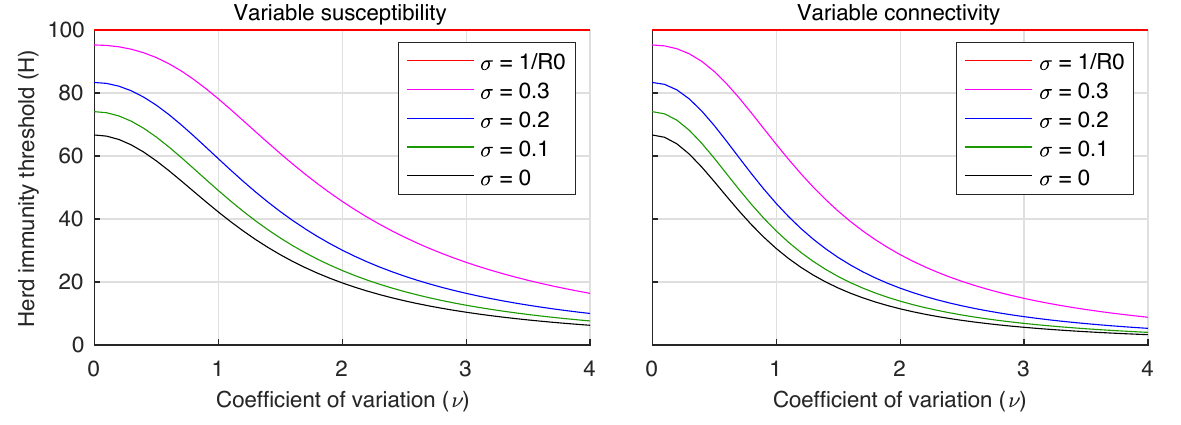}}
\caption{Herd immunity threshold with reinfection. Curves correspond to the SEIR model with reinfection (Eqs. \ref{Sreinf}-\ref{Rreinf}) assuming $\Ro=3$ and gamma distributed susceptibility (left, Eq. \ref{H0si_hetsus}) or connectivity (right, Eq. \ref{H0si_hetcon}). Efficacy of naturally acquired immunity is captured by a reinfection parameter $\sigma$, potentially ranging between $\sigma =0$ (100\% efficacy) and $\sigma =1$ (0 efficacy). Five values of the reinfection parameter are depicted: $\sigma =0$ (black); $\sigma =0.1$ (green); $\sigma =0.2$ (blue); $\sigma =0.3$ (magenta); and $\sigma =1/\Ro$ (red). Above $\sigma =1/\Ro$  (reinfection threshold \cite{Gomes2004,Gomes2016}) the infection becomes stably endemic and there is no herd immunity threshold.}\label{fig:HIT_RT}
\end{figure}

Curves assuming a selection of values for $\sigma$ are represented graphically in Figure \ref{fig:HIT_RT}. Note the critical behaviour at the reinfection threshold ($\sigma=1/\Ro$) in red, which separates the regime where individual immunity is sufficiently potent for the herd immunity threshold to be achievable from the regime where endemicity will establish.

\section{Model with a carrier state}  \label{sec: carrier}

In applications to the COVID-19 pandemic \cite{Gomes2022,Aguas2020} the exposed compartments are not simply a latent state but a carrier state where individuals are infectious but to a lesser degree than those in the fully infectious compartment. Building on the reinfection model (Eqs. \ref{Sreinf}-\ref{Sreinf}) we now introduce parameter $\rho\leq 1$ to denote the ratio of infectiousness between exposed and fully infectious individuals.  
What changes is the force of infection $\lambda$:
\begin{eqnarray*}
\mbox{Variable susceptibility:} 	&\quad& 	\lambda(t) = \b \int \rho E(x,t)+ I(x,t)\ dx = \beta\ (\rho \Esf(t)+ \Isf(t)),   \\
\mbox{Variable connectivity:} 	&\quad& 	\lambda(t) = \b \int x (\rho E(x,t)+ I(x,t))\ dx  =  \beta\ (\rho \Eb(t)+ \Ib(t)). \\
\end{eqnarray*}

The basic and effective reproduction numbers become:
\[
\mbox{Variable susceptibility:} 
\quad 
\Ro = \beta\left(\frac{\rho}{\delta}+\frac{1}{\gamma}\right)
\quad\mbox{and}\quad
{\Reff}(t) = \beta\left(\frac{\rho}{\delta}+\frac{1}{\gamma}\right) (\Sb(t)+\si\Rb(t)) ,
\]
\[
\mbox{Variable connectivity:}
\quad 
\Ro = (1+\nu^2) \beta\left(\frac{\rho}{\delta}+\frac{1}{\gamma}\right)
\quad\mbox{and}\quad
{\Reff}(t) = \beta\left(\frac{\rho}{\delta}+\frac{1}{\gamma}\right) (\Sbb(t)+\si\Rbb(t)).
\]

The calculation of the effective reproduction number ${\Reff}$ is slightly different. 
The difference is that now we have to add the time an individual is incubating the infection in $\Esf$ to the infectious period, multiplied by the factor $\rho$.
Since the average time an individual spends in $\Esf$ is $1/\delta$ we get:

\begin{itemize}
\item $ {\Reff}(t) = \beta (\rho/\delta+1/\gamma)\ (\Sb(t)+ \si \Rb(t))    $ in the variable susceptibility case, and
\item $ {\Reff}(t) =   \beta (\rho/\delta+1/\gamma)\ (\Sbb(t)+ \si \Rbb(t))  $ in the variable connectivity case,
\end{itemize}
where $\Sb(t)$ and $\Sbb(t)$ are the moments of $S(x,t)$ defined above, and the same with $R(x,t)$.

In particular, we get $\Ro =  \beta (\rho/\delta+1/\gamma)$ and $\Ro = (1+\nu^2) \beta (\rho/\delta+1/\gamma)$, respectively, and ${\Reff} =  \Ro\ (\Sb+ \si \Rb)$ and ${\Reff} = \Ro/(1+\nu^2)\ (\Sbb+ \si \Rbb)$.
Finally, we obtain the same formulas for the herd immunity threshold in terms of $\Ro$ and $\nu$ (Section \ref{HIT}, without reinfection) or more generally $\Ro$, $\nu$ and $\sigma$ (Section \ref{sec:reinfection}, with reinfection).

\section{Discussion}

After completion of this work, similar ideas for capturing individual variation with mean-field epidemic models were elaborated \cite{Britton,DiLauro,Kawagoe,Neipel,Rose,Tkachenko}. These recent developments were largely prompted by the COVID-19 pandemic. In agreement with our results, \cite{Rose,Neipel} find that when susceptibility is initially gamma distributed it remains so through the course of the epidemic, leading naturally to power-law behaviour in the force of infection \cite{Novozhilov}. In addition the authors show that other initial distributions converge towards gamma through the process of contagion. Other authors \cite{Kawagoe} derive epidemic final sizes assuming alternative distributions of susceptibility, considering in addition that infectivity may exhibit some correlation with susceptibility (such as in the variable connectivity models analyzed here). They compare numerical results for gamma and lognormal distributions with those obtained using an empirical distribution of individual contacts derived from cell phone geolocation data. Focussing on variable connectivity \cite{Britton,DiLauro} address herd immunity thresholds using age-structured compartmental models. Additionally \cite{DiLauro} consider a variety of non-pharmaceutical intervention scenarios, emphasizing subtle results when interventions change the contact network. Another group of authors \cite{Tkachenko} distinguishes between persistent and transient individual variation to highlight that only the former is subject to the kind of selection that lowers epidemic final sizes and herd immunity thresholds. 

Despite being prompted by COVID-19 none of the above references attempted to quantify the individual variation which was under selection by the force of infection and hence contributed to lower epidemic final sizes and herd immunity thresholds. This is done in our associated work \cite{Gomes2022,Aguas2020}, where the problem is inverted and selectable variation is inferred from its effects on epidemic patterns. Although we and others had previous adopted similar approaches to other infectious diseases \cite{Finkenstadt,Smith,Bellan,Corder} the work has been processed more cautiously during the pandemic due to the greater implications that estimating lower herd immunity thresholds and epidemic final sizes might have for policies and behaviors in this context.

The concept of herd immunity was originally developed in the context of vaccination programs (\cite{Goncalves,Fine}). Defining the percentage of the population that must be immunised to cause infection prevalence to decline, the concept has provided a useful target for vaccination coverage. In hypothetical scenarios of vaccines delivered at random and individuals mixing at random, the herd immunity threshold is given by the simple expression ($1-\Ro^{-1}$) when immunity is fully protective. Concretely, for $\Ro$ between 2.5 and 5, this would indicate that 60-80\% random subjects would need be immunized to prevent spread of infection. This formula would not apply, however, if vaccination programmes were designed to prioritize more connected individuals, for instance \cite{Elbasha}. Similarly, it does not apply when immunity is acquired in response to natural infection, which does not occur at random. Individuals who are more susceptible or more exposed to infection are prone to be infected and become immune earlier. As a result, earlier episodes contribute disproportionately to herd immunity as they remove highly susceptible and exposed subjects from the susceptible pool \cite{Ferrari,Britton,Kawagoe,Neipel,Rose,Gomes2022,Aguas2020}. In our basic models, the herd immunity threshold becomes $\H=1-\Ro^{-1/(1+\nu^2)}$ in the case of gamma distributed susceptibility, and $\H= 1-\Ro^{-1/(1+2\nu^2)}$ with gamma distributed connectivity (exposure), which decline sharply when coefficients of variation ($\nu$) increase from 0 to 2, remaining below 20\% for more variable populations in a particular illustration where $\Ro=3$ (Figure \ref{fig:HIT}). The magnitude of the decline depends on what property is heterogeneous and how it is distributed among individuals, but the downward trend is robust provided that acquired immunity is efficacious enough to keep transmission below the reinfection threshold (Figure \ref{fig:HIT_RT}) ($\sigma<\Ro^{-1}$, where $\sigma$ is the susceptibility of individuals who have recovered relative to their respective susceptibility prior to infection) \cite{Gomes2004,Gomes2016}. In our reinfection models, herd immunity thresholds are derived as $\H=1-\left( (\Ro^{-1}-\si)/(1-\si)\right)^{1/(1+\nu^2)}$ in the case of gamma distributed susceptibility, and $\H= 1-\left( (\Ro^{-1}-\si)/(1-\si)\right)^{1/(1+2\nu^2)}$ with gamma distributed connectivity, when $\sigma<\Ro^{-1}$. If immunity is not potent enough to keep the system below the reinfection threshold then a herd immunity threshold is not attainable and the disease persists in stable endemicity, irrespective of individual variation. 

Finally, we stress that the herd immunity threshold ($\H$) is a theoretical framework to assess epidemic potential to the same extent that $\Ro$ is a theoretical framework. Their interdependence shows that if $\Ro$ increases due to evolution of the infectious agent, for example, so does $\H$. Also, if new susceptibles enter the population through birth or other processes, or if immunity wanes or is evaded by pathogen lineages, a previously acquired herd immunity status may be lost leaving the population prone to subsequent outbreaks. Furthermore, if transmission has a marked seasonal pattern the same level of immunity may place the population above threshold in low season and below threshold in high season, in a cyclical manner. Although $\H$ is not as immediately applicable as often implied, it is a more informative measure of epidemic potential than $\Ro$ given that it accounts for variation in susceptibility or exposure ($\nu$) in addition to average transmissibility $\Ro$. The more accurately we know $\H$ the better we can assess trade-offs and inform public health policy.

\appendix

\numberwithin{equation}{section}
\renewcommand{\theequation}{\thesection.\arabic{equation}}


\section{Derivations of moments $\Si$, $\Sb$ and $\Sbb$ for gamma distributed traits} \label{app:gamma}

In this appendix we derive explicit formulas for $\Sb$ and $\Sbb$, in the cases of variable susceptibility and connectively respectively, in terms of $\Si$ when the traits are gamma distributed.

We use conventional notation for the gamma probability distribution: 
\begin{eqnarray*}
\gam_{\alpha,\beta}(x) &=& \frac{\beta^\alpha}{\Gamma(\alpha)}x^{\alpha-1} e^{-\beta x},
\end{eqnarray*}
noting that $\alpha$ and $\beta$ are the shape and rate parameters, respectively. The notation $\beta$ clashes with the transmission coefficient which is also conventionally denoted by the same Greek letter, but its appearance as rate parameter in this appendix ceases once we condition our gamma distributions to having mean $\alpha/\beta=1$, i.e., $\beta=\alpha$, so confusion can be precluded. 

Given an initial gamma distribution with mean 1, explicitly $q(x) = \gam_{\alpha,\alpha}(x)$, which we substitute in (Eq. \ref{eq: stx}), we get
\begin{eqnarray*}
S(x,t) 	&=& \frac{\alpha^\alpha}{\Gamma(\alpha)}x^{\alpha-1} e^{-x (\alpha+k_t)} \\
		&=& \left(\frac{\alpha}{\alpha+k_t}\right)^\alpha \frac{(\alpha+k_t)^\alpha}{\Gamma(\alpha)}x^{\alpha-1} e^{-x (\alpha+k_t)} \\
&=& \left(\frac{\alpha}{\alpha+k_t}\right)^\alpha \cdot \gam_{\alpha,\alpha+k_t}(x).
\end{eqnarray*}

Using the equalities
$\int  \gam_{\alpha,\alpha+k_t}(x)dx=1$,
$\int x \gam_{\alpha,\alpha+k_t}(x)dx = \alpha/(\alpha+k_t)$ and 
$\int x^2 \gam_{\alpha,\alpha+k_t}(x)dx=[\alpha(\alpha+1)]/[(\alpha+k_t)^2]$,
we calculate $\Si(t)$, $\Sb(t)$ and $\Sbb(t)$:

\begin{eqnarray*}
\Si(t) &=& \left(\frac{\alpha}{\alpha+k_t}\right)^\alpha \cdot \int \gam_{\alpha,\alpha+k_t}(x)dx = \left(\frac{\alpha}{\alpha+k_t}\right)^\alpha,\\
\Sb(t) &=& \left(\frac{\alpha}{\alpha+k_t}\right)^\alpha \cdot \int x \gam_{\alpha,\alpha+k_t}(x)dx = \left(\frac{\alpha}{\alpha+k_t}\right)^\alpha\cdot\frac{\alpha}{\alpha+k_t} = \left(\frac{\alpha}{\alpha+k_t}\right)^{\alpha+1},\\
\Sbb(t) &=&  \left(\frac{\alpha}{\alpha+k_t}\right)^\alpha \cdot \int x^2 \gam_{\alpha,\alpha+k_t}(x)dx = \left(\frac{\alpha}{\alpha+k_t}\right)^\alpha\cdot\frac{\alpha(\alpha+1)}{(\alpha+k_t)^2} = \left(\frac{\alpha}{\alpha+k_t}\right)^{\alpha+2}\cdot \frac{\alpha+1}{\alpha}.
\end{eqnarray*}

\

\noindent From the above we get
\begin{eqnarray}
\Sb(t) &=& \Si(t)^{\frac{\alpha+1}{\alpha}}, \label{Sb_gamma}\\
\Sbb(t) &=& \Si(t)^{\frac{\alpha+2}{\alpha}} \cdot\frac{\alpha+1}{\alpha}. \label{Sbb_gamma}
\end{eqnarray}


\section{Approximate variable connectivity model in $\Si$, $\Ei$, $\Ii$, $\Ri$ variables} \label{app:connectivity}

In (Eqs. \ref{Scon0}-\ref{Rcon0}), we derived a closed system for the variable connectivity model on the variables $\Sb$, $\Eb$, $\Ib$, $\Rb$.
However, we would like to have a closed system on the variables $\Si$, $\Ei$, $\Ii$, $\Ri$, as we did in the variable susceptibility case to enable direct fitting to incidence data (Eqs. \ref{Ssus}-\ref{Rsus}) for parameter estimation and scenario projections.
We propose a system of approximate equations which works well when the infectious period is small as in acute infectious diseases (Eqs. \ref{Scon2}-\ref{Rcon2} below).

The justification for the approximation is based on the assumption that the following two quantities are very close to each other when the infectious period is small compared to the length of the epidemic.
At each time $t$ we consider the following two factors, that we denote $\infe(t)$ and $\inft(t)$
\begin{itemize}
\item $\infe(t)$ is the average infectivity of a typical individual who {\em becomes infected} at time $t$. 
In other words, this is the average value of trait $x$ taken over the pool of individuals that go from $\Ssf$ to $\Esf$ at time $t$.			\label{inf1}
\item $\inft(t)$ is the average infectivity of a typical individual who {\em is infectious} at time $t$. 
In other words, this is the average value of trait $x$ taken over the pool of individuals that are in compartment $\Isf$ at time $t$.			\label{inf2}
\end{itemize}
Both $\infe$ and $\inft$ are formulated below (Eq. \ref{fs}).
The distinction between them is only meaningful in models where infectivity varies among individuals in such a way that those infected in different days tend to have different infectivities --- this occurs in our variable connectivity model (as infectivity is positively correlated with susceptibility, the selection which reduces susceptibility in the susceptible pool over time also reduces infectivity) but not in the case of variable susceptibility (where all individuals have the same infectivity).

\begin{figure}
 \centerline{\includegraphics[clip, trim=5cm 0cm 6cm 0cm, width=0.9\textwidth]{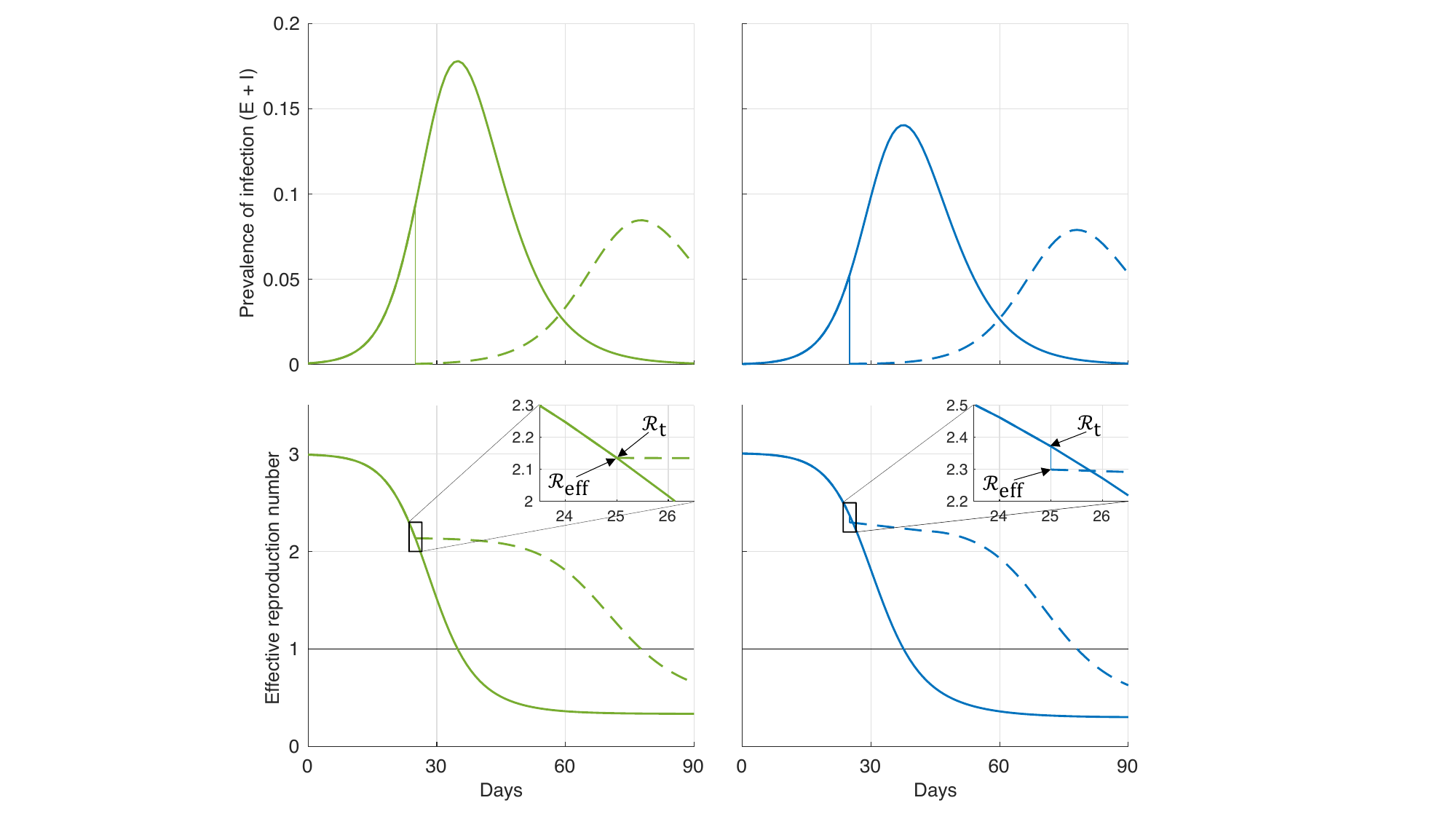}}
\caption{Effective reproduction number. Top panels show epidemic curves generated by running system (Eqs. \ref{introSgen}-\ref{introRgen}) with explicit gamma distributed susceptibility (left) or connectivity (right). Solid curves are unmodified epidemics while dashed show the outcome of hypothetically clearing all infections instantaneously on day 25 and progressing with a small infectious seed. Bottom panels show the respective $\Rt$ values, calculated as $(\beta/\gamma)\cdot \Sb$ in the case of variable susceptibility and $(\beta/\gamma)\cdot (\Ib/\Ii)\cdot \Sb$ in the case of variable connectivity, with or without the intervention on day 25. The empirical index $\Rt$ coincides with the effective reproduction number $\Reff$ in the variable susceptibility case but is slightly higher when variation is in connectivity. This is highlighted by zooming into the areas marked by black rectangles (see insets). Parameters: $\Ro=3$; $\delta=1/4$ per day; $\gamma=1/3$ per day; and $\nu=1$.}\label{fig:Reffs}
\end{figure}

Note that we use the factor $\infe$ in our calculation of $\Reff$ in (Eq. \ref{Reff_con}), where we multiply $\beta$ by the average infectivity $\infe$ (same as average connectivity), the average susceptibility $\Sb$, and the length of the infectious period $(1/\gamma)$:
\begin{eqnarray}
\Reff&=&\frac{\beta}{\gamma}\cdot \infe \cdot \Sb. \label{Reff_f}
\end{eqnarray}
Let us also note that the ``$R$-number'' which was popularized in the COVID-19 pandemic, loosely described as the number of new infections caused by a typical individual who is infectious at time $t$, is defined in terms of $\inft$ rather than $\infe$. This more empirical notion, formulated as $\Rt= (\beta/\gamma)\cdot \inft \cdot \Sb$, is not always coincident with the basic reproduction number $\Reff$, introduced in textbooks as the spectral radius of the next generation operator \cite{DHB} and adopted in this paper. The two notions differ when the average infectivity of individuals infected in a particular day is different to the average infectivity of those who have been infected in another day. However, this difference is of little consequence to variable connectivity models when the infectious period is small and of no consequence when variation is in susceptibility (see Figure \ref{fig:Reffs}).

We claim that the values of $\infe$ and $\inft$ are given by the following formulas:
\begin{eqnarray}
\infe = \Sbb/\Sb \quad &\mbox{ and }& \quad \inft=\Ib/ \Ii.    \label{fs}
\end{eqnarray}
The second equality follows from the formula for average infectivity, $\int (x I(x,t)/ \Ii) dx$, for every time $t$. The first equality was essentially derived in Section \ref{Reff} when we calculated $\Reff$ for variable connectivity (Eq. \ref{Reff_con}).
Within that calculation we had to consider the infectivity of an individual who becomes infected at time $t$, which is equal to $\int x p(x,t) dx$, where $p(x,t)$ is the  density function measuring the probability that such an individual has connectivity level $x$.
We saw that since $p(x,t)$ is proportional to $x S(x,t)$, we have that $p(x,t) = x S(x,t)/\Sb(t)$.
It follows then that $\infe(t) = \int x^2 S(x,t)/\Sb(t) dx$, for every $t$, which is the expression above.

We now can write our assumption as:
\begin{eqnarray}
{\Sbb}/{\Sb}  &\approx& {\Ib}/{\Ii}.  \label{approx}
\end{eqnarray}

With this we can derive the approximate closed system of equations on the variables $\Si$, $\Ei$, $\Ii$, $\Ri$.
Observe that if we integrate the equations (\ref{introSgen})-(\ref{introRgen}) over $x$ and use that $\lambda = \beta \ \Ib$, we obtain 
\begin{eqnarray}
\frac{d\Ssf}{dt} &=& - \beta\ \Ib\ \Sb\\
\frac{d\Esf}{dt}  &=& \beta\ \Ib\ \Sb -\delta\ \Esf \\ 
\frac{d\Isf}{dt}  &=&  \delta\ \Esf - \gamma\ \Isf\\
\frac{d\Rsf}{dt} &=&  \gamma\ \Isf. 
\end{eqnarray}
Then, using that $\Ib\ \Sb  \approx \Ii\ \Sbb$ (Eq. \ref{approx}), and that with gamma distributed traits we have $\Sbb =  (1+\nu^2)\ \Si(t)^{1+2 \nu^2}$ (Eq. \ref{Sbarbar}), we get the approximate closed system we seek:

\begin{eqnarray}
\frac{d\Ssf}{dt} &\approx&  - (1+\nu^2)\ \beta\ \Isf\ \Ssf^{1+2\nu^2} , \label{Scon2}\\ 
\frac{d\Esf}{dt} &\approx&  (1+\nu^2)\ \beta\ \Isf\ \Ssf^{1+2\nu^2}  -\delta\ \Esf, \label{Econ2}\\ 
\frac{d\Isf}{dt}  &= & \delta\ \Esf - \gamma\ \Isf, \label{Icon2}\\
\frac{d\Rsf}{dt}  &= & \gamma\ \Isf. \label{Rcon2}
\end{eqnarray}

\begin{figure}
 \centerline{\includegraphics[clip, trim=3.5cm 9cm 3.5cm 7cm, width=0.6\textwidth]{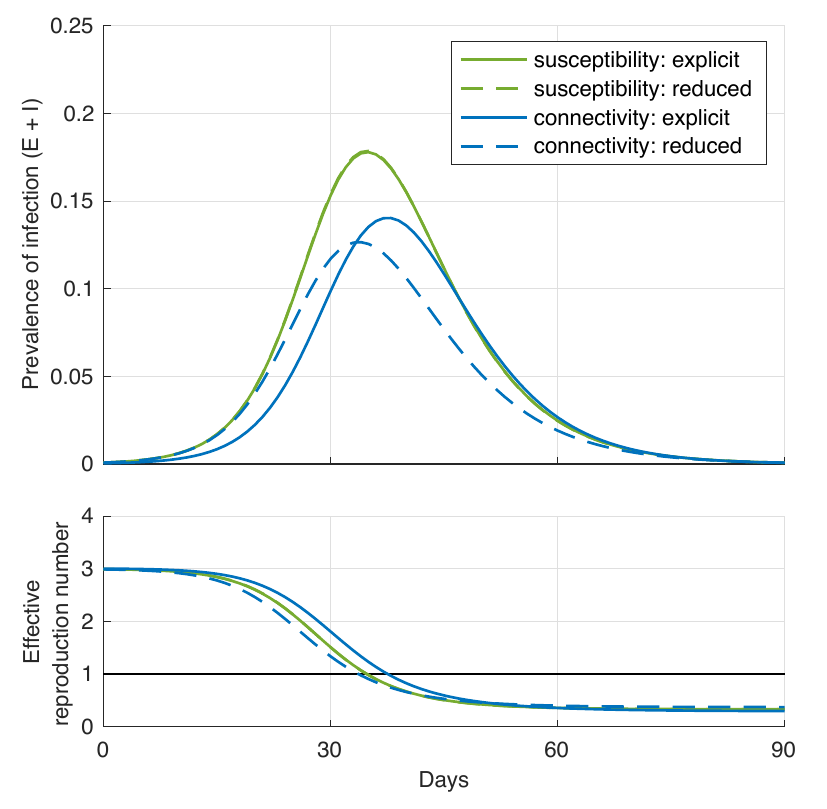}}
\caption{Epidemics with gamma distributed susceptibility or connectivity. Solid curves were generated by running a discretized version of the infinite system (Eqs. \ref{introSgen}-\ref{introRgen}) with explicit gamma distributed susceptibility (green) or connectivity (blue). Dashes curves are the corresponding outputs of the reduced systems (Eqs. \ref{Ssus}-\ref{Rsus}) and (Eqs. \ref{Scon2}-\ref{Rcon2}). Prevalences of infection ($\Esf+\Isf$) over time are depicted on the top panel and effective reproduction numbers 
on the bottom. Parameters: $\Ro=3$; $\delta=1/4$ per day; $\gamma=1/3$ per day; and $\nu=1$.}\label{fig:epidemic}
\end{figure}

Figure \ref{fig:epidemic} illustrates epidemic curves generated with system (Eqs. \ref{introSgen}-\ref{introRgen}) with an explicit gamma distribution (mean 1 and coefficient of variation $\nu=1$) for susceptibility (solid green) and connectivity (solid blue). Technically the distribution was discretized in $n$ bins and a system of $4 n$ ODEs was run to generate the curves. Dashed curves were added to the plots to enable comparisons with the corresponding results generated by the reduced systems (Eqs. \ref{Ssus}-\ref{Rsus} and \ref{Scon2}-\ref{Rcon2}, respectively). The bottom panel shown the associated effective reproduction numbers $\Reff$
in the case of the reduced systems, and the empirical proxy $\Rt$ when explicit distributions are implemented (recall that the two coincide when variation is in susceptibility). All runs were conducted with basic reproduction number $\Ro=3$, incubation period ($1/\delta=4$ days), infectious period ($1/\gamma=3$ days) and coefficient of individual variation ($\nu=1$). Notice that the reduction is exact in the variable susceptibility case but not with variable connectivity. In a related study we fitted both explicit and reduced models to real epidemics and obtained similar parameter estimates and conclusions irrespective of which version was adopted.

\section*{Acknowledgements}

We thank Paul McKeigue and Joel Miller for valuable discussions and two anonymous reviewers for constructive criticisms and suggestions.


\begin{thebibliography}{xx}
\bibitem{Aguas2020}  {\sc R. Aguas}, {\sc G. Gon\c calves}, {\sc M. U. Ferreira}, {\sc M. G. M. Gomes}, Herd immunity thresholds for SARS-CoV-2 estimated from unfolding epidemics. {\em medRxiv} 10.1101/2020.07.23.20160762v4.
\bibitem{AM1991} {\sc R. M. Anderson}, {\sc R. M. May}, {\em Infectious Diseases of Humans: Dynamics and Control}.  Oxford University Press Inc., New York (1991).
\bibitem{AMMJ} {\sc R. M. Anderson}, {\sc G. F. Medley}, {\sc R. M. May}, {\sc A. M. Johnson}, A preliminary study of the transmission dynamics of the human immunodeficiency virus (HIV), the causative agent of AIDS. {\em IMA J Math. Appl. Med. Biol.} 3, 229-263 (1986).
\bibitem{Ball} {\sc F. Ball}, Deterministic and stochastic epidemic models with several kinds of susceptibles. {\em Adv. Appl. Probab.} {\bf 17}, 1-22 (1985).
\bibitem{Bellan} {\sc S. E. Bellan}, {\sc J. Dushoff}, {\sc A. P. Galvani}, {\sc L. A. Meyers}, Reassessment of HIV-1 Acute Phase Infectivity: Accounting for Heterogeneity and Study Design with Simulated Cohorts. {\em PLOS Med.} 12, e1001801 (2015).
\bibitem{Britton} {\sc T. Britton}, {\sc F. Ball}, {\sc P. A. Trapman},  A mathematical model reveals the influence of population heterogeneity on herd immunity to SARS-CoV-2. {\em Science} 369, 846-849 (2020).
\bibitem{Corder} {\sc R. M. Corder}, {\sc M. U. Ferreira}, {\sc M. G. M. Gomes}, Modelling the epidemiology of residual Plasmodium vivax malaria in a heterogeneous host population: a case study in the Amazon Basin. {\em PLOS Comput. Biol.} {\bf 16}, e1007377 (2020).
\bibitem{DiLauro} {\sc F. Di Lauro}, {\sc L. Berthouze}, {\sc M. D. Dorey}, {\sc J. C. Miller}, {\sc I. Z. Kiss},  The Impact of Contact Structure and Mixing on Control Measures and Disease-Induced Herd Immunity in Epidemic Models: A Mean-Field Model Perspective. {\em Bull. Math. Biol.} 83, 1-25 (2021).
\bibitem{DHB} {\sc O. Diekmann}, {\sc J. A. P. Heesterbeek}, {\sc T. Britton},  {\em Mathematical tools for Understanding Infectious Disease Dinamics.} Princeton University Press, Princeton, New Jersey (2013).
\bibitem{Elbasha} {\sc E. H. Elbasha}, {\sc A. B. Gumel}, Vaccination and herd immunity thresholds in heterogeneous populations. {\em J. Math. Biol.} 83, 239 (2021).
\bibitem{Ferrari} {\sc M. J. Ferrari}, {\sc S. Bansal}, {\sc L. A. Meyers}, {\sc O. N. Bjornstad}, Network frailty and the geometry of herd immunity. {\em Proc. R. Soc.} B 273, 2743-2748 (2006).
\bibitem{Finkenstadt} {\sc B. F. Finkenstadt}, {\sc B. T. Grenfell}, Time series modelling of childhood diseases: a dynamical systems approach. {\em Appl. Statist.} {\bf 49}, 187-205 (2000).
\bibitem{Fine} {\sc P. Fine}, {\sc K. Eames}, {\sc D. L. Heymann}, “Herd immunity”: a rough guide, {\em Clin. Infect. Dis.} 52, 911-916 (2011).
\bibitem{Gart68} {\sc J. J. Gart}, The mathematical analysis of an epidemic with two kinds of susceptibles. {\em Biometrics} {\bf 24}, 557-566 (1968).
\bibitem{Gart71} {\sc J. J. Gart}, The statistical analysis of chain-binomial epidemic models with several kinds of susceptibles. {\em Biometrics} {\bf 28}, 921-930 (1971).
\bibitem{Gomes2022} {\sc M. G. M. Gomes}, {\sc M. U. Ferreira}, {\sc R. M. Corder}, {\sc J. G. King}, {\sc C. Souto-Maior}, {\sc C. Penha-Gon\c calves}, {\sc G. Gon\c calves}, {\sc M. Chikina}, {\sc W. Pegden}, {\sc R. Aguas}, Individual variation in susceptibility or exposure to SARS-CoV-2 lowers the herd immunity threshold. {\em J. Theor. Biol.} 540, 111063 (2022).
\bibitem{Gomes2016} {\sc M. G. M. Gomes}, {\sc E. Gjini}, {\sc J. S. Lopes}, {\sc C. Souto-Maior}, {\sc C. Rebelo}, A theoretical framework to identify invariant thresholds in infectious disease epidemiology. {\em J. Theor. Biol.} 395, 97-102 (2016).
\bibitem{Gomes2004} {\sc M. G. M. Gomes}, {\sc L. J. White}, {\sc G. F. Medley}, Infection, reinfection, and vaccination under suboptimal immune protection: Epidemiological perspectives. {\em J. Theor. Biol.} 228, 539-549 (2004).
\bibitem{Goncalves} {\sc G. Gon\c calves}, Herd immunity: recent uses in vaccine assessment. {\em Expert Rev. Vaccines} 7, 1493-1506 (2008).
\bibitem{Heesterbeek}  {\sc H. Heesterbeek}, et al., Modeling infectious disease dynamics in the complex landscape of global health. {\em Science} 347, aaa4339 (2015).
\bibitem{Kawagoe}  {\sc K. Kawagoe}, {\sc M. Rychnovsky}, {\sc S. Chang}, {\sc G. Huber}, {\sc L. M. Li}, {\sc J. Miller}, {\sc R. Pnini}, {\sc B. Veytsman}, {\sc D. Yllanes}, Epidemic dynamics in inhomogeneous populations and the role of superspreaders. {\em Phys. Rev. Research} 3, 033283 (2021).
\bibitem{KM}  {\sc W. O. Kermack}, {\sc A. G. McKendrick}, A contribution to the mathematical theory of epidemics. {\em Proc. R. Soc. Lond.} A 115, 700-721 (1927).
\bibitem{McKendrick} {\sc A. G. McKendrick}, The dynamics of crowd infection. {\em Edinb. Med. J.} {\bf 47}, 117–136 (1939).
\bibitem{Miller} {\sc J. C. Miller}, {\sc A. C. Slim}, {\sc E. M. Volz}, Edge-based compartmental modelling for infectious disease spread. {\em J. R. Soc. Interface} {\bf 9}, 890-906 (2012).
\bibitem{Neipel} {\sc J. Neipel}, {\sc J. Bauermann}, {\sc S. Bo}, {\sc T. Harmon}, {\sc F. J\"ulicher},  Power-Law population heterogeneity governs epidemic waves. {\em PLOS One} 15, e0239678 (2020).
\bibitem{Novozhilov}  {\sc A. S.	Novozhilov}, On the spread of epidemics in a closed heterogeneous population. {\em Math. Biosci.} 215, 177–185 (2008).
\bibitem{P-S&V} {\sc R. Pastor-Satorras}, {\sc A. Vespignani}, Epidemic dynamics and endemic states in complex networks. {\em Phys. Rev. E} {\bf 63}, 066117 (2001).
\bibitem{Rose}  {\sc C. Rose}, {\sc A. J. Medford}, {\sc C. F. Goldsmith}, {\sc T. Vegge}, {\sc J. S. Weitz}, {\sc A. A. Peterson}, Heterogeneity in susceptibility dictates the order of epidemic models. {\em J. Theor. Biol.} 528, 110839 (2021).
\bibitem{Ross}  {\sc R. Ross}, An application of the theory of probabilities to the study of a priori pathometry, Part I. {\em Philos. Trans. R. Soc. Lond. A} 92, 204–230 (1916).
\bibitem{RH}  {\sc R. Ross}, {\sc H. H. Hudson}, An application of the theory of probabilities to the study of a priori pathometry, Part II. {\em Philos. Trans. R. Soc. Lond. A} 93, 212–225 (1917).
\bibitem{Smith}  {\sc D. L. Smith}, {\sc J. Dushoff}, {\sc R. W. Snow}, {\sc S. I. Hay}, The entomological inoculation rate and its relation to the prevalence of Plasmodium falciparum infection in African children. {\em Nature} {\bf 438}, 492-495 (2005).
\bibitem{Tkachenko} {\sc A. V. Tkachenko}, {\sc S. Maslov}, {\sc A. Elbanna}, {\sc G. N. Wong}, {\sc Z. J. Weiner}, {\sc N. Goldenfeld},  Time-dependent heterogeneity leads to transient suppression of the COVID-19 epidemic, not herd immunity. {\em Proc. Natl. Acad. Sci. U. S. A.} 118, e2015972118 (2021).
\bibitem{Woolhouse} {\sc M. E. J. Woolhouse}, {\sc C. Dye}, {\sc J.-F. Etard}, {\sc T. Smith}, {\sc J. D. Charlwood}, {\sc G. P. Garnett}, {\sc P. Hagan}, {\sc J. L. K. Hii}, {\sc P. D. Ndhlovu}, {\sc R. J. Quinnell}, {\sc C. H. Watts}, {\sc S. K. Chandiwana}, {\sc R. M. Anderson},  Heterogeneities in the transmission of infectious agents: Implications for the design of control programs. {\em Proc. Natl. Acad. Sci. U. S. A.} 94, 338-342 (1997).


\end{thebibliography}
\end{document}